\documentclass{WileyMSP-template}
\usepackage{graphicx,amssymb,amsmath}
\usepackage{color}
\usepackage[dvipsnames]{xcolor}
\begin{document}

\pagestyle{fancy}
\rhead{\includegraphics[width=2.5cm]{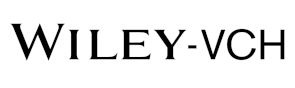}}

\title{Statistical-Mechanical Theory on the Probability Distribution\\ Function for the Net Charge of an Electrolyte Droplet}

\maketitle

\author{Yuki Uematsu}
\author{Keiju Suda}

%\dedication{Optional dedication here. If no dedication is required, please leave blank}
\dedication{}

\begin{affiliations}
Y. Uematsu\\
Department of Physics and Information Technology, Kyushu Institute of Technology, Iizuka 820-8502, Japan\\
PRESTO, Japan Science and Technology Agency, 4-1-8 Honcho, Kawaguchi, Saitama 332-0012, Japan\\
Email Address: uematsu@phys.kyutech.ac.jp

K. Suda\\
Department of Physics and Information Technology, Kyushu Institute of Technology, Iizuka 820-8502, Japan\\

\end{affiliations}


%\date{\today}
%\bibliographystyle{elsarticle-num}

\begin{abstract}
Droplets of electrolyte solutions in an insulating medium are ubiquitous in nature. 
The net charges of these droplets are normally nonzero, and they fluctuate.    
However, a theory on the probability distribution function for the net charge of droplets is lacking, so far. 
Thus, a statistical-mechanical theory of a charged droplet is developed with including the effect of the electrostatic energy of charging as well as the random distribution of ions. 
Then, the probability distribution function for the net charge of an electrolyte droplet is calculated assuming that it is generated from a macroscopic solution with the different cation and anion concentrations.
Using the Gaussian approximation and Stirling's formula, the analytic results for the average and variance of the net charge of a droplet are obtained.
\end{abstract}

\section{Introduction}

Droplets of electrolyte solutions in an insulating medium play a crucial role in many physico-chemical processes such as cloud droplets \cite{Twomey1956,Takahashi1973,Harrison2008, Burgo2015}, electro- and sonic spray \cite{Beauchamp2002, Jarrold2006,Zlich2008, Jarrold2011, Colussi2006, Enami2010, Mishra2019, Beauchamp2015, Gordon2008}, and water-in-oil emulsions and droplets \cite{Tsao_1998,Tsao_1999, Uematsu_2022, Kang2008,Esmaeilzadeh2010,Kang2011,Esmaeilzadeh2012,Choi2013,Mishra2020, Burgo2020, Bishop2018,Kurimura2013, Oldham2002, Zare2020, Consta2019, Consta2020, Consta2021}.
The enhancement of the chemical reactivity in charged droplets is one of the most debated topics recently \cite{Ingram_2016, Qiu_2022, Mehrgardi_2022, Colussi_2023, Nguyen_2023}. 
The size of these droplets is typically from $100\,$nm to $1\,$mm, and the droplets are usually charged. 
The simplest way to generate relatively large droplets is using a capillary with control of the capillary radius, liquid flux, and imposed pressure \cite{Choi2013,Mishra2020,Artemov_2023}. 
To obtain smaller size of droplets, aerodynamic breakups \cite{Wierzba_1990} and/or Rayleigh instability of a charged droplet after evaporation \cite{Fenn2000} were often used.
The droplets in an insulating liquid are, similarly, generated by a high-shear homogenizer, ultrasonic irradiation \cite{Li_1978,Li_1978_2}, or a well-designed microfluidic device such as flow-focusing geometry \cite{Shah_2008}. 
In each process, the droplets are possibly charged, and the generation of charged droplets is considered to be a complicated combination of hydrodynamics \cite{Zlich2008,Jarrold2011}, electrostatics \cite{Mishra2019, Artemov_2023, Fenn2000}, and physical chemistry of interface \cite{Colussi2006,Enami2010, Choi2013,Mishra2020}.
However, the quantitative prediction of the net charge of a droplet is still a challenging problem.
Because the experimental data of droplets has a distribution of their net charge \cite{Zlich2008}, a statistical-mechanical point of view is also important.
However, no statistical-mechanical theory was constructed except for the random charge distribution \cite{Zlich2008,Jarrold2011}.

Choi {\it et al.}~studied the spontaneous electrical charging of millimeter-sized aqueous droplets generated by conventional pipetting \cite{Choi2013}, and they obtained a charge of about $+10^9 e$ for the water droplet with a radius of about $1\,$mm, where $e$ is the elementary charge. 
The detected charge was proportional to the surface area of the pipette tip inside, and the dependency on the salinity and the pH had a similar trend to the electrokinetic surface charge density of the tip surface inside. 
Thus, they interpreted that the charge of the droplet originates from the charge in a diffuse layer of a negatively charged tip surface. 
After that, Mishra {\it et al.}~found that if the pipette is filled with water and connected to a water reservoir the droplet charge becomes negative \cite{Mishra2020}.
Furthermore, Artemov {\it et al.}~ found that the charge of $0.1\,$mm-sized droplets is linear with the electrostatic potential of the solution, and its capacitance is determined by the droplet size \cite{Artemov_2023}.
When the solution is electrically grounded, the droplet charge, however, remains nonzero and has an order of $10^6e$.
The sign of the charge depends on the material of the tip surface and pH of the solution \cite{Artemov_2023}.  

Jarrold {\it et al.}~studied electric charges of micrometer-sized water droplets generated in the aerodynamic breakups of parental droplets  \cite{Zlich2008}.
They measured the charge distributions of the droplets generated by sonic spray and a vibrating orifice aerosol generator. 
Even though the droplet generation mechanism is not electrically biased, their charge distributions are biased toward positive, and the average and square root of the variance are both the order of $10^4 e$ for the droplets with a $2.7\,\mu$m radius.
They interpreted the biased positive charge as a consequence of the fact that a parental droplet was deformed to be a positively charged annulus and a negatively charged thin film, and the detected micrometer-sized droplets were generated from the positively charged annulus \cite{Zlich2008}.
The reason for the negatively charged film is that the air/water interface is normally negatively charged \cite{Uematsu_2021}.
This picture was further confirmed by the findings that the film droplets generated by a bubble rapture at the air/water interface were negatively charged \cite{Jarrold2011}, and the ion-specific effect on the film droplets detected by the mass spectrometry \cite{Enami_2013}.

These studies revealed that understanding of the charging mechanisms remains qualitative. 
Thus, it is necessary to study the quantitative prediction of the charge amount of a droplet.
In this paper, we develop a statistical-mechanical theory of the probability distribution function for the net charge of an electrolyte droplet in an insulating medium.
The electrostatic energy of the droplet is taken into account in addition to the translational entropy of the ions.
We use the grand-canonical ensemble to calculate the probability distribution function for the net charge, and the average and variance are evaluated numerically or using approximations.

\section{Model}

\begin{figure}
\center
\includegraphics[width=4.5cm]{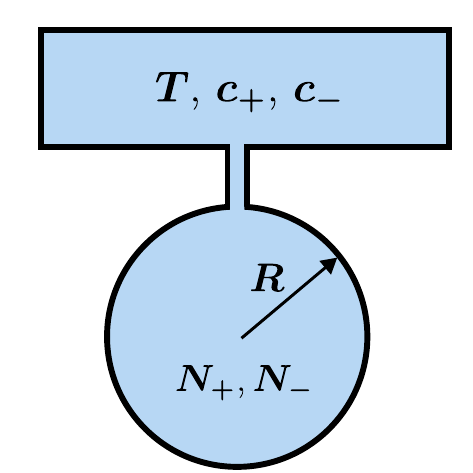}
\caption{
Illustration of the model.
The reservoir is characterized by the temperature $T$, the cation concentration $c_+$, and the anion concentration $c_-$.
The grand canonical ensemble yields the probability distribution function for the cation and anion numbers $N_+$ and $N_-$ included in the droplet with radius $R$. 
}
\label{fig.1}
\end{figure}

Considering an electrolyte droplet with the radius $R$ in an insulating medium generated from the macroscopic region of the electrolyte solution which has the cation concentration $c_+$ and $c_-$ at the temperature $T$ as illustrated in Fig.~\ref{fig.1}. 
The number of cations inside the droplet is set by $N_+$, and the number of anions by $N_-$, then, according to the canonical ensemble, the partition function is given by
\begin{equation}
Z(N_+,N_-) = \frac{V^{N_+}\lambda_+^{-3N_+}}{N_+!}\frac{V^{N_-}\lambda_-^{-3N_-}}{N_-!}\exp\left(-\frac{\ell_\mathrm{0}(\Delta N)^2}{2R}\right)
\end{equation}
where $V=4\pi R^3/3$ is the volume of the droplet, $\lambda_+$ and $\lambda_-$ are the thermal de Broglie lengths of cations and anions, $\ell_\mathrm{0}=e^2/4\pi\varepsilon_0 k_\mathrm{B}T$ is the vacuum Bjerrum length, $\varepsilon_0$ is the vacuum permittivity, $k_\mathrm{B}$ is the Boltzmann constant, and $\Delta N=N_+-N_-$ is the net charge divided by $e$. 
Here, the electrostatic field is assumed to be spherically symmetric and to vanish far away from the droplet.
Then, the grand partition function is 
\begin{equation}
\begin{split}
\Xi & = \sum_{N_+=0}^\infty\sum_{N_-=0}^\infty Z(N_+,N_-)\,\mathrm{e}^{\frac{\mu_+N_++\mu_-N_-}{k_\mathrm{B}T}} \\
& = \sum_{N_+=0}^\infty\sum_{N_-=0}^\infty \frac{(c_+V)^{N_+}(c_-V)^{N_-}}{N_+!N_-!}\mathrm{e}^{-\ell_\mathrm{0}(\Delta N)^2/2R} \\
%& \approx \int^\infty_{-\infty} dN_+\int^\infty_{-\infty} dN_- \frac{\mathrm{exp}\left[-\frac{(N_+-c_+^0V)^2}{2c_+^0V}-\frac{(N_--c_+^0V)^2}{2c_-^0V}+(c_+^0+c_-^0)V-\ell_\mathrm{0}(\Delta N)^2/2R\right]}{2\pi V\sqrt{c_+^0c_-^0}}
\end{split}
\label{eq:gpf}
\end{equation}
where $\mu_+ = k_\mathrm{B}T\ln(c_+ \lambda_+^3)$ and $\mu_-=k_\mathrm{B}T\ln(c_-\lambda_-^3)$ are the chemical potentials of cations and anions in the reservoir.

To obtain the probability distribution function for the net charge, we transform the variables from $N_+$ and $N_-$ to $\bar{N}=(N_++N_-)/2$ and $\Delta N$ and sum up the grand partition function concerning $\bar N$. 
Then, we obtain
\begin{equation}
\begin{split}
&\Xi = \\
&\sum_{\Delta N=-\infty}^\infty\sum_{\bar N=\frac{|\Delta N|}{2}}^{\infty} \frac{(c_+V)^{\bar N+\frac{\Delta N}{2}}(c_-V)^{\bar N-\frac{\Delta N}{2}}}{(\bar N+\frac{\Delta N}{2})!(\bar N-\frac{\Delta N}{2})!}\mathrm{e}^{-\ell_\mathrm{0}(\Delta N)^2/2R},\\
\end{split}
\label{eq:xi2} 
\end{equation}
where $\sum_{\bar N=\frac{|\Delta N|}{2}}^{\infty}$ means the sum of $\bar N = \frac{|\Delta N|}{2}, \frac{|\Delta N|}{2}+1, \frac{|\Delta N|}{2}+2, \cdots$ and $\sum_{\Delta N =-\infty}^\infty$ denotes the sum in the order of $\Delta N=0$, $\pm 1$, $\pm 2$, $\cdots$.
Therefore, the probability distribution function of the net charge of a droplet is obtained as
\begin{equation}
P(\Delta N) = \frac{\mathrm{e}^{-\frac{\ell_\mathrm{0}}{2R}(\Delta N)^2}}{\Xi}\sum_{\bar N=\frac{|\Delta N|}{2}}^{\infty} \frac{(c_+V)^{\bar N+\frac{\Delta N}{2}}(c_-V)^{\bar N-\frac{\Delta N}{2}}}{(\bar N+\frac{\Delta N}{2})!(\bar N-\frac{\Delta N}{2})!}.
\label{eq:prob}
\end{equation}
However, it is impossible to analytically evaluate eqs.~\ref{eq:xi2} and \ref{eq:prob}.

\section{Results and Discussion}

\subsection{Gaussian approximation}
To obtain the analytic equation for the probability distribution function, we approximate the part of the Poisson distribution by the Gaussian distribution. 
This approximation needs the condition of $c_i V\gg 1$. 
Then, the Poisson distribution in eq.~\ref{eq:gpf} can be approximated by the Gaussian distribution as
\begin{equation}
\frac{(c_iV)^{N_i}}{N_i!}\approx \frac{\mathrm{exp}\left[-\frac{(N_i-c_iV)^2}{2c_iV}+c_iV\right]}{\sqrt{2\pi c_iV}} \textrm{ for } i=\pm,
\end{equation} 
where we assume that the space of the stochastic variables $N_+$ and $N_-$ is continuous and $(-\infty,\infty)\otimes (-\infty,\infty)$. 
Because the Jacobian $|\partial(N_+,N_-)/\partial(\bar{N},\Delta N)| = 1$, 
\begin{equation}
\begin{split}
\sum_{N_+=0}^\infty \sum_{N_-=0}^\infty \cdots & \approx \int^\infty_{-\infty} \int^{\infty}_{-\infty} \cdots dN_+dN_- \\
& =\int^\infty_{-\infty} \int^{\infty}_{-\infty} \cdots d\bar Nd(\Delta N). 
\end{split}
\end{equation}
The integral of $\bar N$ is analytically given by
\begin{equation}
\begin{split}
&\int  \mathrm{exp}\left[-\frac{(N_+-c_+V)^2}{2c_+V}-\frac{(N_--c_-V)^2}{2c_-V}\right] \frac{d\bar{N}}{\sqrt{(2\pi V)^2c_+c_-}}\\
&= \frac{1}{\sqrt{2\pi (c_++c_-)V}}\exp\left[-\frac{(\Delta N-(c_+-c_-)V)^2}{2(c_++c_-)V}\right].
\end{split}
\end{equation}
Therefore, the grand partition function is obtained as 
\begin{equation}
%\Xi =  C\int^\infty_{-\infty} \mathrm{e}^{-\frac{\ell_0}{2R}(\Delta N)^2-\frac{(\Delta N-(c_+-c_-)V)^2}{2(c_++c_-)V}} d(\Delta N),
\Xi =  \frac{\displaystyle \mathrm{e}^{(c_++c_-)V}\int^\infty_{-\infty} \mathrm{e}^{-\frac{\ell_0}{2R}(\Delta N)^2-\frac{(\Delta N-(c_+-c_-)V)^2}{2(c_++c_-)V}} d(\Delta N)}{\sqrt{2\pi(c_++c_-)V}}.
\end{equation}
The probability distribution function for $\Delta N$ is obtained as
\begin{equation}
P(\Delta N) =  \frac{\displaystyle \mathrm{e}^{-\frac{\ell_0}{2R}(\Delta N)^2-\frac{(\Delta N-(c_+-c_-)V)^2}{2(c_++c_-)V}} }{\displaystyle \int^\infty_{-\infty} \mathrm{e}^{-\frac{\ell_0}{2R}(\Delta N)^2-\frac{(\Delta N-(c_+-c_-)V)^2}{2(c_++c_-)V}} d(\Delta N)}.
\label{eq:pdf}
\end{equation}

Eq.~\ref{eq:pdf} is the product of the two Gaussian distributions with different averages and variances, and it is also a Gaussian distribution itself. 
The average of $\Delta N$ is given by
\begin{equation}
\langle \Delta N \rangle = \frac{(c_+ -c_- )V}{1+(\ell_0/R)(c_+ +c_- )V},
\label{eq:mean}
\end{equation}
and the variance of $\Delta N$ is given by
\begin{equation}
\langle (\delta\Delta N)^2 \rangle = \frac{(c_+ +c_- )V}{1+(\ell_0/R)(c_+ +c_- )V},
\label{eq:std}
\end{equation}
where $\delta \Delta N = \Delta N - \langle\Delta N\rangle$.
Eqs.~\ref{eq:mean} and \ref{eq:std} imply that if $(\ell_0/R)(c_+ +c_- )V\ll 1$, the ion partition to the droplet is just random as given by $\langle \Delta N \rangle = (c_+ -c_- )V$ and $\langle (\delta\Delta N)^2 \rangle = (c_+ +c_- )V$. 
Otherwise, the ion partition is affected by the electrostatic energy of the droplet.  

\begin{figure}
\center
\includegraphics{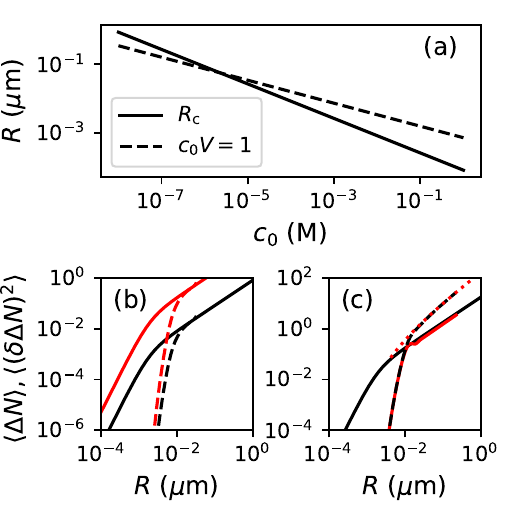}
\caption{
(a). The solid line denotes $R=R_\mathrm{c}$ as a function of a given salt concentration $c_0$, whereas the broken line is $c_0 V = 1$. 
(b). $\langle \Delta N \rangle$ and $\langle (\delta \Delta N)^2 \rangle$ as a function of $R$ with $c_+=1.1\,$mM and $c_- =1\,$mM. 
The black solid and broken lines denote $\langle \Delta N \rangle$ calculated by the Gaussian approximation and the finite series given by  eqs.~\ref{eq:xi2} and \ref{eq:prob}.
The red solid and broken lines denote $\langle (\delta \Delta N)^2 \rangle$ calculated by the Gaussian approximation and the finite series given by  eqs.~\ref{eq:xi2} and \ref{eq:prob}.
(c). $\langle \Delta N \rangle$ and $\langle (\delta \Delta N)^2 \rangle$ as a function of $R$ with $c_+=1\,$mM and $c_- =0\,$mM. 
The black solid and broken lines denote $\langle N_+ \rangle$ calculated by the Gaussian approximation and the finite series given by eq.~\ref{eq:gpf}.
The red dotted line is $\langle \Delta N \rangle$ calculated by eq.~\ref{eq:root} whereas the red solid line $\langle (\delta \Delta N)^2 \rangle$ calculated by the finite series given by  eq.~\ref{eq:gpf}.
}
\label{fig:02}
\end{figure}

To examine whether the Gaussian approximation works, we define the critical radius $R_\mathrm{c}$ at a given concentration $c_0$ by
\begin{equation}
R_\mathrm{c}=\sqrt{\frac{3}{4\pi\ell_0 c_0}}.
\end{equation}
This $R_\mathrm{c}$ determines whether the net charge distribution is just random ($R\ll R_\mathrm{c}$) or affected by the electrostatics $(R\gg R_\mathrm{c})$. 
In Fig.~\ref{fig:02}a, the black line denotes the critical radius $R_\mathrm{c}$ as a function of $c_0$, whereas the red line denotes $c_0 V=1$. 
To use the Gaussian approximation, the condition $c_iV\gg 1$ is needed.  
However, considering the realistic salt concentration of water ($c_0\ge 10^{-7}\,$M), almost all of the region of $c_0V \gg 1$ satisfies $R \gg R_\mathrm{c}$, suggesting that the net charge distribution is not random but affected by the electrostatics. 

To discuss more detail,  we set $c_+=1.1c_0=1.1\,$mM and $c_- = c_0=1\,$mM. 
Fig.~\ref{fig:02}b shows the average and variance of $\Delta N$ as a function of $R$ with fixed $c_0 = 10^{-3}\,$M in a double-logarithmic scale.  
The black solid line is the average $\langle \Delta N \rangle$ using Gaussian approximation (eq.~\ref{eq:mean}), and the black broken line is the exact average calculated by eqs.~\ref{eq:xi2} and \ref{eq:prob} with summing up of $\bar N$ from $|\Delta N|/2$ to $|\Delta N|/2+70$ and $\Delta N$ from $0$ to $\pm 70$.
The black broken line exhibits a slope steeper than that of the black solid line in small $R$, and the black broken line agrees with the black solid line in $R>(3/4\pi c_0)^{1/3}$.
The red solid line is the variance $\langle (\Delta N)^2 \rangle$ using Gaussian approximation (eq.~\ref{eq:std}), and the red broken line is the exact average calculated as the black broken line.
Again, the red broken line exhibits a slope steeper than that of the red solid line in small $R$, and it agrees with the red solid line in $R>(3/4\pi c_0)^{1/3}$.
Therefore, we conclude that Gaussian approximation works well for $R>R_\mathrm{c}$ and $R>(3/4\pi c_0)^{1/3}$. 
However, for $R<(3/4\pi c_0)^{1/3}$, the Gaussian approximation cannot predict the average and variance of the charge of the droplet because the number of ions inside the droplet is relatively less. 

\subsection{Beyond Gaussian Approximation}

Secondly, we consider the case $c_+ \gg c_-$. 
Such a situation is ubiquitous, for example, droplet generation from already charged parental droplets.
%Because water interfaces are usually charged with the surface potential, typically about $-50\,$mV. 
%This yields the ratio of the concentrations in the diffuse layer as $c_+/c_- = \mathrm{e}^{-2e\psi_0/k_\mathrm{B}T} \approx 55(\gg 1)$.
For simplicity, we consider the case $c_+=c_0$ and $c_-=0\,$mM. 
The average of the net charge using the Gaussian approximation is given by
\begin{equation}
\begin{split}
\langle \Delta N \rangle & = \frac{c_0V}{1+(\ell_\mathrm0/R)c_0V}= \left\{\begin{array}{ll}
c_0V & \textrm{ for } R \ll R_\mathrm{c} \\
R/\ell_0 & \textrm{ for } R \gg R_\mathrm{c} \\
\end{array}
\right.. 
\end{split}
\label{eq:mean2}
\end{equation}
This result is intriguing. 
When $R \ll R_\mathrm{c}$, the average of the net charge is the product of the concentration in the reservoir and the volume.  
However, $R \gg R_\mathrm{c}$, the mean of the net charge is $R/\ell_0$, which is independent of $c_0$.
The variance of the net charge for the case $c_-=0\,$mM in the Gaussian approximation is given by $\langle (\delta\Delta N)^2 \rangle  = \langle \Delta N \rangle$. 

In Fig.~\ref{fig:02}c, the black solid line is the average $\langle \Delta N\rangle$ when $c_+=c_0=1\,$mM and $c_-=0\,$mM using Gaussian approximation. 
The black broken line is the exact sum up to $N_+=70$, which significantly deviates from the black solid line in the entire range of $R$. 
Thus, for the case $c_+ \gg c_-$, the Gaussian approximation does not work.
To evaluate $\langle \Delta N \rangle$ analytically, we approximate the factorial term $(N_+!)\approx (N_+/\mathrm{e})^N$ by Stirling's formula.
Then, the probability for the case $c_-=0\,$mM is 
\begin{equation}
\begin{split}
P(\Delta N) & = \frac{1}{\Xi}\frac{(c_+ V)^{\Delta N}}{(\Delta N)!}\mathrm{e}^{-\ell_0(\Delta N)^2/2R} \\
& \approx \frac{1}{\Xi}\mathrm{e}^{\Delta N\ln(c_+ V)-\Delta N\ln(\Delta N)+\Delta N-\frac{\ell_0}{2R}(\Delta N)^2}. 
\end{split}
\end{equation}   
When $\langle \Delta N \rangle \gg \sqrt{\langle (\Delta N)^2 \rangle }$, the average is determined by the position of the maximum of the probability distribution function.
Thus, the root of the equation 
\begin{equation}
\frac{d}{d(\Delta N)}(-\ln P(\Delta N)) = 0,
\label{eq:root}
\end{equation}
is the average $\langle \Delta N \rangle$. 
The red dotted line in Fig.~\ref{fig:02}c is the root of eq.~\ref{eq:root}, which agrees with the exact sum up to $N_+ =70$ (black broken line) for $R>(3/4\pi c_0)^{1/3}(=7\,\mathrm{nm})$.
The red solid line is the variance calculated by exact sum up to $N_+=70$, and it agrees with the Gaussian approximation (black solid line) for $R>(3/4\pi c_0)^{1/3}$ except for the region around $R=10^{-2}\,\mu$m.

\subsection{Discussion about the applicability of the theory}

Our theory corresponds to a situation where a single droplet is generated from a macroscopic reservoir with the ion concentrations $c_+$ and $c_-$.
Furthermore, the electrostatic energy of the reservoir after the generation of a droplet is neglected.  
This is rationalized when the reservoir size is much larger than the droplet size (water-filled case in Ref.~\cite{Mishra2020}) or the reservoir is grounded by inserting a metallic electrode into the reservoir solution \cite{Artemov_2023}. 
Although the reservoir is electrically neutral in these experiments \cite{Mishra2020,Artemov_2023}, the average of the net charge is nonzero probably because the specific ion adsorption to the air/water or tip/water interface.
These effects, however, are not included in our theory, and thus, the direct comparison between our theory and the experimental data is difficult. 
In addition, these experiments did not provide the probability distribution function of the net charge of the droplet, and therefore, the discussion of the variance is also difficult. 
The direct comparison with experimental studies needs more data on the probability distribution function of the droplet charge. 

\begin{figure}
\center
\includegraphics{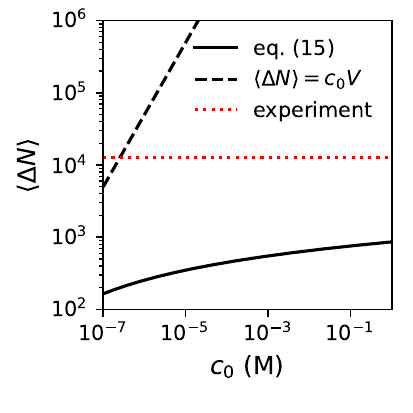}
\caption{
(a). The black solid line denotes the average net charge $\langle \Delta N \rangle$ calculated by eq.~\ref{eq:root} as a function of $c_0$ with $R=2.7\,\mu$m, $c_+=c_0$, and $c_-=0\,$mM. 
The black broken line denotes $\langle \Delta N \rangle=c_0 V$.
The red dotted line is the experimental value of the average net charge $\langle \Delta N \rangle = 1.26\times 10^4$ for pure water with $R=2.7\,\mu$m \cite{Zlich2008}.
}
\label{fig:03}
\end{figure}

Another process to generate droplets is the split of a single droplet into many equally sized droplets.
Our theory cannot apply to this situation because the sum of their charges should be equal to the charge of the original single droplet. 
Furthermore, in practice, the sizes of many droplets generated from a single droplet have a broad distribution \cite{Kolmogorov_1941,Gorokhovski_2003}.
When the mean charge of a droplet is a function of the size, the size distribution significantly affects the charge distribution. 
To see the applicability in more detail, we discuss the previous experimental data on the droplets generated by sonic spray \cite{Zlich2008}.
As mentioned in the introduction, this study measured the charge distribution function of the droplets generated from the aerodynamic breakup of positively charged annulus parts in the parental droplets. 
Here, we try to compare the experimental results with our theory.

Fig.~\ref{fig:03}a shows the average net charge of the droplet as a function of $c_0$ with the conditions of $R=2.7\,\mu$m, $c_+=c_0$, and $c_-=0\,$mM.
Because $R$ is much larger than $R_\mathrm{c}$, we calculate $\langle \Delta N \rangle$ by solving eq.~\ref{eq:root} and denote it by the black solid line.
The black broken line denotes $\langle \Delta N \rangle=c_0 V$, which is the result of random distribution neglecting the electrostatic energy.
In the experiments, pure water ($c_0=10^{-7}\,$M) was used, and the droplet was considered to be generated from positively charged annulus parts in the parental droplets.
Therefore, we compare the experimental results with our theory of $c_+=c_0$ and $c_-=0\,$mM. 
The red dotted line is the experimental value for the droplet charge $\langle \Delta N\rangle = 1.26\times 10^4$ \cite{Zlich2008}.
It is close to the result of random distribution neglecting the electrostatic energy (black broken line) rather than considering the electrostatic energy (black solid line). 
This can be explained as follows.
For simplicity, equally sized two-droplet generation from a single positively charged droplet containing $N_0(=c_0\cdot 2V)$ cations is considered.
When the numbers of cations in the two droplets are $N_1$ and $N_2$, the conservation of the cations yields $N_0=N_1+N_2$. 
If the electrostatic energy of both droplets is taken into account, it is proportional to $(N_1)^2 + (N_0-N_1)^2$. 
Thus, the minimum energy principle gives the average charge of $N_0/2(=c_0V)$, which seems to be neglecting the electrostatic effect. 
We think that this mechanism dominates in multiple-droplet generation from a single charged droplet.

\section{Conclusion}

We developed the statistical-mechanical theory on the probability distribution function for the net charge of an electrolyte droplet. 
In the theory, the suppression of the net charge induced by the electrostatic energy is included as well as the random distribution of the ions.
Using the formulation of a grand canonical ensemble, the probability distribution function of the net charge, eq.~\ref{eq:prob}, is obtained.
Although the obtained probability distribution includes an infinite series which cannot be calculated analytically, we approximate it by using the Gaussian approximation or Stirling's formula and obtain the analytic equation for the probability for the net charge.  
Furthermore, the theoretical results are discussed along with the recent experimental studies \cite{Zlich2008,Mishra2020,Artemov_2023}.

Our theory does not consider the ion distribution inside the droplet nor the structure of the double layer \cite{Uematsu_2022}.
Furthermore,  the specific ion adsorption significantly affects the charge separation of the droplet \cite{Enami_2013, Enami_2014}.
Additionally, in a sub-micrometer droplet, the average ion separation, the Debye length, and Bjerrum length, and the droplet size are all in the same order \cite{Enami_2013, Enami_2014}.  
Therefore, it is necessary to seriously consider the ion distribution inside the droplet when we apply the theory to more complicated experiments \cite{Choi2013,Mishra2020,Artemov_2023}. 
However, we believe that this theory is the first step to predicting and controlling the charge of droplets and emulsions in insulating media from the statistical-mechanical point of view. 

\medskip
\textbf{Acknowledgements} \par %delete if not applicable))

This work was supported by JST PRESTO (Grant No. JPMJPR21O2) and a grant from the Kao Foundation for Arts and Sciences. 

\medskip

\bibliographystyle{MSP}
\bibliography{droplet}

\begin{figure}
\textbf{Table of Contents}\\
\medskip
\center
  \includegraphics{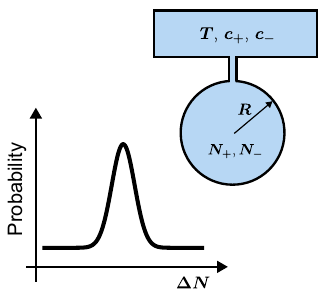}
  \medskip
  \caption*{A statistical-mechanical theory of a charged droplet is developed with including the effect of the electrostatic energy of charging as well as the random distribution of ions.
Then, the probability distribution function for the net charge of an electrolyte droplet is calculated assuming that it is generated from a macroscopic solution with the different cation and anion concentrations.}
\end{figure}

\end{document}